\documentstyle[prl,aps,epsf]{revtex}

\begin{document}

\title{Reconstruction of the $\nu =1$ Quantum Hall Edge}
\author{A. Karlhede and K. Lejnell}

\address{
Department of Physics,
Stockholm University,
Box 6730, S-11385 Stockholm,
Sweden
}

\maketitle
\begin{abstract}
The sharp $\nu=1$ quantum Hall edge present for hard confinement
is shown to have two modes that go soft as the confining potential softens. This 
signals a second order transition to a reconstructed edge that  is either
a depolarized spin-texture edge or a polarized charge density wave edge. 
\end{abstract}



\section{Introduction}

In the Quantum Hall (QH) effect at ferromagnetic fillings there are 
excitations ``skyrmions'' that involve spin textures - topologically 
nontrivial configurations of the spins.\cite{sondhi1} There is  
experimental evidence that skyrmions are 
the lowest energy charged excitations at $\nu=1$.\cite{barrett1,schmeller,goldberg} 
A basic feature 
of these spin textures is that the topological density 
is proportional to the charge density. As a consequence, low energy smooth 
variations in the charge density can be achieved by texturing the spins. 
We here discuss another example where this mechanism seems to  be at work, 
namely at the edges of QH systems.\cite{karlhede,karlhede2,franco} 

We consider the edge of a $\nu=1$ QH system as the confining potential softens from 
a hard confinement where the edge is sharp. Calculating the excitations about the sharp
edge we find two modes that soften as the confining potential softens. One is a spin flip mode
and one is a polarized mode. The softening modes signal second order phase transitions to 
a spin texture edge and a charge density wave edge respectively. For small 
Zeeman energies the initial instability is to the spin textured edge. 
We predict that the textured edge has a new gapless mode. 
Experimentally, the textured edge is signalled by its sensitivity to the value 
of the Zeeman energy and the associated depolarization of the edge.

\section{The Sharp Edge}

Our system is  a two-dimensional electron gas in a perpendicular magnetic 
field $\bf B$. The electron gas is confined to a wide bar with 
two straight 
edges. We assume the confining potential at the left edge always to be strong so 
that the system is a spin polarized $\nu=1$ quantum Hall state at this edge and 
that this state continues deep into the bulk. The orbital Hilbert 
space is restricted to the lowest Landau level.
 
We consider the right edge of this  $\nu=1$ quantum Hall liquid as a function of 
the strength of the confining potential at this edge. When the potential is 
strong  the ground state is a polarized $\nu=1$ state where all the spin up 
orbitals from $k=0$ (left edge) out to a maximum momentum $k_F$ (right edge)
are filled:
\begin{equation}
|\nu =1\rangle = \prod _{0\leq k \leq k_F} c^{\dagger}_{k \uparrow} |0\rangle \ \ \ .
 \label{nu1}
\end{equation}
$c^\dagger _{k\sigma}$ creates electrons  in the lowest Landau level 
with  momentum $k=2\pi n_k /L$, $n_k=0,\pm 1,\pm 2, ...$ and 
spin $\sigma=\uparrow, \downarrow$. The corresponding 
wave functions are $\varphi_k=(\sqrt{\pi} L \ell)^{-1/2}e^{iky}
e^{-(x/\ell-k\ell)^2/2}$, where $\ell=\sqrt{\hbar c/eB}$ is the magnetic length. 
$L$ is the length of the edge and we assume 
periodic boundary conditions along the edge and use 
Landau gauge ${\bf A}=Bx \hat {\bf y}$. 

The state $|\nu=1\rangle$ 
has a {\it sharp} spin polarized edge: In momentum space the density falls 
discontinuously to zero, 
whereas in real space it falls to zero over a length of order $\ell$. This is the
simplest of the sharp quantum Hall edges that have been studied extensively, see 
e.g. \cite{wenreview}.

The Hamiltonian is constructed by taking matrix elements of the Coulomb 
interaction $V({\bf r})= e^2/(\epsilon |{\bf r}|)$ between states in the 
lowest Landau level, in the presence of a background charge density $\rho_b 
({\bf r})$ which makes the system neutral and confines the electron gas. 
For now, we follow standard practise and implement the confining potential by 
taking  $\rho_b$ to fall linearly
from a constant bulk value $(2\pi \ell^2)^{-1}$ to $0$ over a distance $w$ at the 
edge. However, it turns out that this gives a 
non-generic confining potential, see below. 
The Hamiltonian also contains the Zeeman 
term $H_{\cal Z}= g\mu_B BS_z$, where $S_z$ is the component of the total spin 
along $\bf B$. The problem is characterized by two dimensionless 
parameters: $\tilde g= g\mu_B B/(e^2/\epsilon \ell)$, the ratio of the Zeeman 
energy ($g\mu_B B$) to the typical Coulomb energy ($e^2/\epsilon \ell$) and 
$\tilde w = w/\ell$, which is a measure of the ``softness'' of the edge.

We study the edge and the edge modes as functions of $\tilde g$ and $\tilde w$.
When $\tilde w =0$, the sharp edge ($|\nu=1\rangle$) is the ground state for any 
$\tilde g$. However, when $\tilde w$ increases, charge will eventually move outwards 
and the edge reconstructs.\cite{macdonald1,chamon,reconstruct} The question is 
how this happens. It is also clear that for large enough $\tilde g$ the 
ground state is spin polarized.

We first consider particle-hole excitations of the sharp edge $|\nu=1\rangle$. 
Excitations of 
this ferromagnetic ground state are characterized by two quantum numbers:
momentum $q$ and spin $s$, corresponding to translations along the edge 
and rotations of the spins about the $z$-axis respectively. 
The possible ph-excitations are 
\begin{eqnarray} \label{polexc}
|q,s=0\rangle &=& \sum_{k_F-q< k\leq k_F} \psi _{k\uparrow}
c^\dagger_{k+q\uparrow}c_{k\uparrow} |\nu=1\rangle \ \ \  
({\rm polarized \, excitations})  \\
|q,s=1 \rangle &=& \sum_{0\leq k\leq k_F} \psi _{k\downarrow}
c^\dagger_{k+q\downarrow}c_{k\uparrow} |\nu=1\rangle \ \ \  
({\rm spin \, flip \, excitations})  \nonumber \ \ \ .
\end{eqnarray}
The wave functions $\psi_{k\sigma}$ and the energies are determined by diagonalizing
the Hamiltonian in the ph-subspace.
\begin{figure}[htbp]
  \begin{center}
  \leavevmode
    \epsfxsize = 9.5cm
    \epsfbox{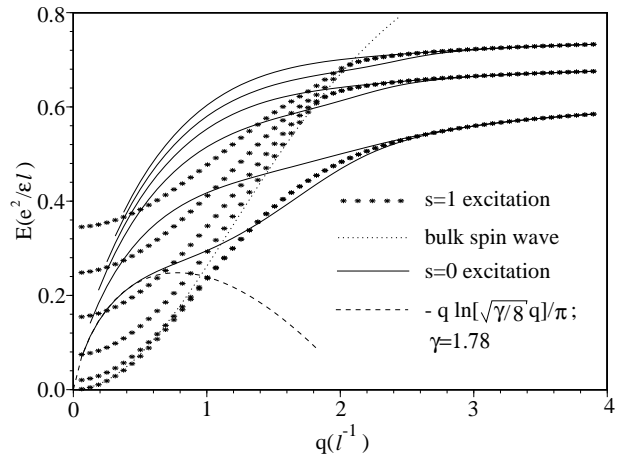}
  \end{center}
\caption{Excitations of energy $E$ and momentum $q$ of the sharp edge $|\nu =1\rangle$ at 
$\tilde w = 0$ and $\tilde g = 0$.}
\label{exc0}
\end{figure}
Fig. \ref{exc0} shows
the lowest energy excitations at $\tilde w=0$ and $\tilde g=0$ for $q \geq 0$. 
The $s=0$ modes are the 
gapless chiral
($q\geq 0$) edge magnetoplasmons, corresponding to 
one-particle, two-particle etc. branches of the one dimensional massless field 
theory.\cite{wenreview} The lowest branch agrees with the analytic result 
$E=-\frac q \pi \ln (\sqrt{\gamma/8}q) , \, \gamma\approx 1.78$.\cite{volkov} 
The $s=1$ excitations are 
non-chiral and the  gapless mode is the 
Goldstone mode of the quantum Hall ferromagnet. The higher energy modes all have 
gaps. For small $q$  these branches are above the bulk spin wave energy, 
but at some critical momentum, $q_c$, each branch falls 
below this energy. For $q<q_c$, the states extend into the bulk, whereas they 
become localized at the edge for $q>q_c$.
The Zeeman energy is included, $\tilde g \neq 0$, by shifting all $s=1$ energies by $\tilde g$.
(The pairing of branches seen in the figure is due to the two edges of the QH bar.)

\begin{figure}[htbp]
  \begin{center}
  \leavevmode
    \epsfxsize = 9.5cm
    \epsfbox{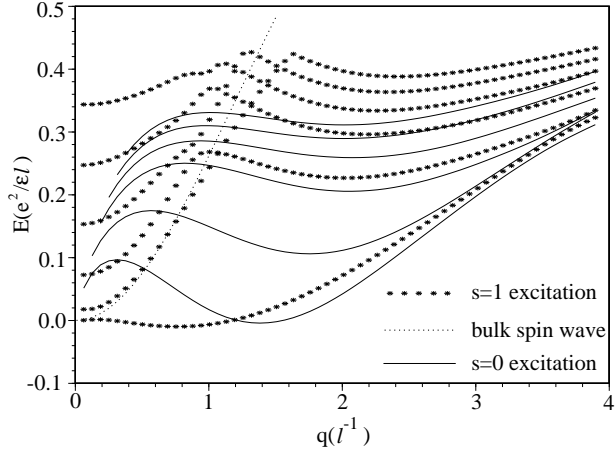}
  \end{center}
\caption{Excitations of energy $E$ and momentum $q$ of the sharp edge $|\nu =1\rangle$ at  
$\tilde w = 7.04$ and $\tilde g = 0$.}
\label{exc7}
\end{figure}

Upon softening the confining potential, i.e., increasing $\tilde w$ the lowest $s=0$ and 
$s=1$ 
modes soften and for each mode the energy becomes negative at some 
critical $\tilde w_c$ (and for some $q$) thus signalling an 
instability of the sharp edge, see Fig. \ref{exc7}. 
At vanishing Zeeman energy, $\tilde g =0$, 
this happens first for the spin flip mode at
$\tilde w_{sf} = 6.77$ and subsequently, at $\tilde w_{pol} = 7.29$, for 
the polarized mode. A nonzero $\tilde g$ disfavours 
the spin flip mode and there is a critical $\tilde g_c$ above which 
the spin flip instability is preempted by the 
polarized instability. (For the $\tilde w$-model $\tilde g_c=0.016$, but for a more realistic
confining potential $\tilde g_c$ is likely to be larger, see below.) The polarized reconstruction 
where a lump of quantum Hall liquid is split off and deposited a distance 
$\ell$ from the bulk happens only at $\tilde w_{hole}=9.0$.\cite{macdonald1,chamon} 
Thus this instability is preempted by the instabilities 
discussed here.

\section{Textured Edge and CDW Edge}

Having identified two modes that go soft as the confinement softens thus 
signalling instabilities of the sharp edge, we here identify what the 
corresponding new ground states are. The ground state corresponding to the 
spin flip mode is a spin texture state
\begin{equation}
|{\rm TEX},q\rangle = \prod_{0\leq k\leq k_F} (u_k c^\dagger _{k\uparrow}
+v_k c^\dagger _{k+q\downarrow})|0\rangle  \ \ \ ,
\label{texgs}
\end{equation}
where $|u_k|^2+|v_k|^2=1$ ($u_k,v_k$ can be choosen real and positive). 
The sharp edge is obtained if $u_k=1$ for all $k$.
For small deviations from the sharp edge $v_k$ are small and 
$u_k\approx 1$, the ground state then becomes 
\begin{equation}
|{\rm TEX},q\rangle \approx (1+\sum_{0\leq k\leq k_F} v_k c^\dagger _{k+q\downarrow}
c _{k\uparrow})|\nu=1\rangle \ \ \ .
\label{sfcond}
\end{equation}
The textured ground state (\ref{texgs}) can thus be thought of as 
a condensation of spin flip excitations, cf. (\ref{polexc}). 
The state $|{\rm TEX},q\rangle$ is a spin texture state of the same type as the one 
that describes skyrmion quasi particles with topological charge $q$.\cite{fertig}
 
The edge spin texture can also be analyzed within the  nonlinear 
$\sigma-$model (where the spin is represented as a unit vector ${\bf n}({\bf r})$)
that describes the
long distance spin dynamics of a QH ferromagnet.\cite{sondhi1}
The edge spin texture takes the form
\begin{equation}
n_x + i n_y =\sqrt{1-f^2} \, {\rm e} ^{i(ky+\theta)} \ \ , \ \ \ n_z=f(x) \ \ \ ,
\label{ntex}
\end{equation}
where $f=f(x)$ approaches $1$ deep in the bulk and falls below $1$ at the edge. This ansatz leads to 
the topological (and hence charge) density $q({\bf r})=(-k/4\pi)df/dx$. Thus the spin is polarized 
deep in the bulk ($n_z=f=1$), but starts tilting as the edge is approached ($n_z=f<1$). 
Moving along the edge,
the spin in the edge region rotates around the $z$-direction with wave vector $k$. By numerically 
integrating the equations of motion obtained from the non-linear $\sigma-$model one determines $f(x)$ and $k$.
This gives results, for $\tilde g \ll 1$, that agrees with Hartree-Fock calculations for the transition to the
textured edge.  

The ground state that corresponds to the polarized ($s=0$) excitations in
(\ref{polexc}) 
is obtained by replacing $\downarrow$ by $\uparrow$ in (\ref{texgs}) 
(this also restricts the range of $k$)
\begin{equation}
|{\rm CDW},q\rangle = \prod_{k_F-q < k\leq k_F} (u_k c^\dagger _{k\uparrow}
+v_k c^\dagger _{k+q\uparrow})|0\rangle  \ \ \ .
\label{polgs}
\end{equation}
In this state charge is moved outwards at the price of modulating the charge density along the 
edge, thus forming a charge density wave (CDW) edge.

\begin{figure}[htbp]
  \begin{center}
  \leavevmode
    \epsfxsize = 9.5cm
    \epsfbox{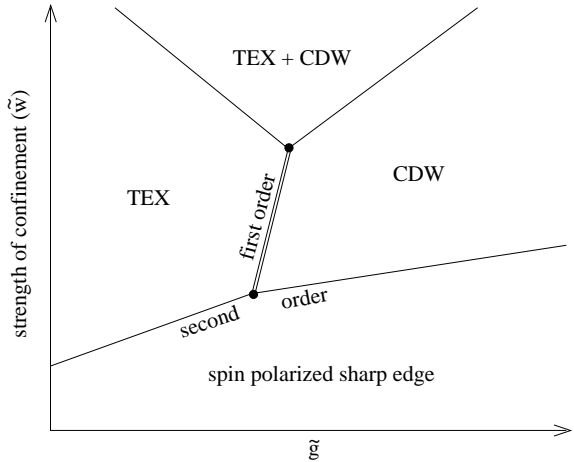}
  \end{center}
\caption{Phase diagram for the $\nu=1$ edge.}
\label{phase}
\end{figure}
Fig. \ref{phase} shows a typical phase diagram in the $\tilde w
\tilde g$-plane. $\tilde w$ should be understood as some measure of a 
confining potential. It turns out that only the topology of the phase diagram is
stable under changes in the confining potential, whereas the  position of the 
phase boundaries (and hence $\tilde g_c$) is  very sensitive. In particular, one finds that
the slope of the softening $s=1$ dispersion curve vanishes at zero momentum  for the 
$\tilde w$-model. Generically, the slope will be negative when the potential softens and this 
will favour the spin flip instability (increasing $\tilde g_c$) since the $s=1$ mode goes 
as $q^2$ whereas the $s=0$ mode goes as $q \ln q$. It is also possible to have a combination of 
a spin texture and a charge density wave edge as indicated in Fig. \ref{phase}.\cite{franco}

\section{Excitations of the Textured Edge}

The sharp edge, $|\nu =1\rangle$, is invariant under translations along the edge, $t_y$, 
as well as under rotations of the spins around the $z$-axis, $s_z$. The textured edge, in Hartree-Fock 
(\ref{texgs}) or in the effective theory (\ref{ntex}), is invariant only under the 
linear combination $t_y+q s_z$, whereas the orthogonal combination is spontaneously broken. 
(In (\ref{ntex}), the angle $\theta$ labels the degenerate ground states and the degenerate 
microscopic states are obtained from (\ref{texgs}) by making a 
rotation of the spins around the $z$-axis.) As a consequence of the broken symmetry there is a
gapless Goldstone mode. Note that the broken symmetry is a symmetry also in the presence of the Zeeman 
term, thus the mode is gapless also for $\tilde g \neq 0$.
When quantum fluctuations are included the broken symmetry will be restored. However, we expect an 
algebraic long range order and that the gapless mode survives.

By first determining the textured ground state, at some point ($\tilde w, \tilde g$),
in Hartree-Fock and then considering ph-excitations,
corresponding to (\ref{polexc}), we obtain the excitations of the textured edge.  
\begin{figure}[htbp]
  \begin{center}
  \leavevmode
    \epsfxsize = 9.5cm
    \epsfbox{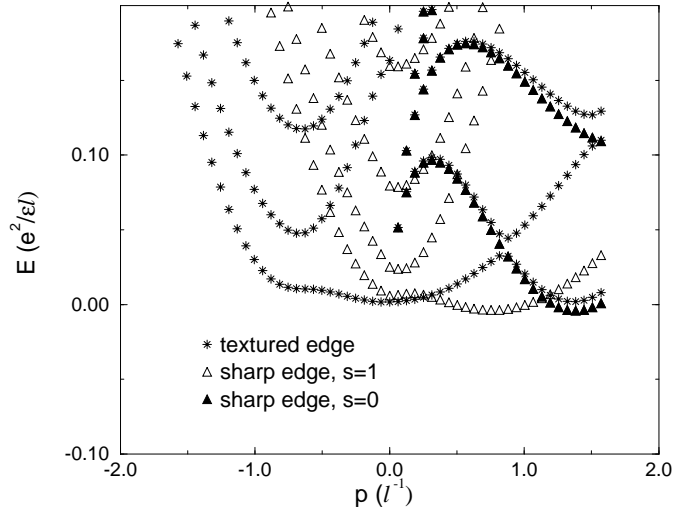}
  \end{center}
\caption{ 
Excitations of energy $E$ and momentum $p$ of  the textured edge at $\tilde w = 7.0, \, \tilde g= 0.006$ and 
$q=0.75$.}
\label{texexc}
\end{figure}
Fig. \ref{texexc} shows the result for $\tilde w=7.04, \, \tilde g=0.006$ and $q=0.75$. 
This is close to the transition from the sharp 
edge and the modes for the sharp edge have been included for comparison. (The data for the sharp 
edge is also for $\tilde w=7.04, \, \tilde g=0.006$; it is not the ground state at this point.) 
We see that the polarized
modes  evolve smoothly into new modes of the textured edge, whereas  
the lowest spin flip mode becomes 
gapless and has a very flat dispersion relation. This mode is concentrated at the edge of the 
system where $v_k \neq 0$. The higher spin flip modes have their minimum translated to 
$p=-0.75$. 

\section{Discussion}

It is believed that standard quantum Hall samples may contain highly reconstructed edges 
with compressible and incompressible regions even for the integer quantum Hall 
states.\cite{multichannels} If this is correct then 
to see the edge reconstructions to a textured (or a charge density wave) edge it may be necessary to 
have a sharper edge, possibly produced by cleaving.\cite{cleave}

The main experimental signature of the textured edge is likely to be that it is depolarized and that
it depends strongly on the Zeeman energy. It could be investigated by tunneling 
into the edge at various
values of a tilted magnetic field or by using NMR.

The edge reconstructions discussed here for the $\nu=1$ edge may take place also at other
ferromagnetic filling factors as well as in quantum dots.\cite{we,oaknin}

\acknowledgements
This talk is based on work with S. A. Kivelson and S. L. Sondhi.
We are also grateful to S. Girvin, T. H. Hansson and A. H. MacDonald 
for useful discussions. The work is supported 
by the  Swedish Natural Science Research Council.

\end{document}